\documentclass[a4paper,twocolumn,superscriptaddress,11pt,accepted=2018-09-14]{quantumarticle}
\pdfoutput=1
\usepackage[utf8]{inputenc}
\usepackage[english]{babel}
\usepackage[T1]{fontenc}
\usepackage{amsmath}
\usepackage{hyperref}

\usepackage{tikz}
\usepackage{lipsum}

\usepackage{graphicx}
\usepackage{dcolumn}
\usepackage{bm}
\usepackage{hyperref}
\usepackage[mathlines]{lineno}

\usepackage{amsmath,amssymb}  
\usepackage{dsfont}
\usepackage{mathrsfs}
\usepackage{amsfonts}
\usepackage{dsfont}
\usepackage{IEEEtrantools}
\usepackage{amsthm}

\usepackage[numbers,sort&compress]{natbib}
\usepackage{stmaryrd}
\usepackage{relsize}

\usepackage{csquotes}
\usepackage{enumerate}
\usepackage{pgfplots}
\usetikzlibrary{calc,positioning}
\usepackage{xcolor}
\usepackage{dsfont}

\usepackage{IEEEtrantools}

\definecolor{qcol}{RGB}{82,41,125}

\newtheorem{definition}{Definition}
\newtheorem{Postulate}{Postulate}

\renewcommand{\H}{\mathcal{H}}
\newcommand{\ket}[1]{|#1\rangle}
\newcommand{\bra}[1]{\langle#1|}
\newcommand{\braket}[2]{\langle#1|#2\rangle}

\newcommand{\proj}[1]{|#1\rangle\langle#1|}

\newcommand{\da}{\downarrow}
\newcommand{\ua}{\uparrow}

\DeclareMathOperator{\tr}{Tr}

\begin{document}

\title{On Formalisms and Interpretations}
\date{\today}
\author{Veronika Baumann}
\affiliation{Faculty of Informatics, Universit\`a della Svizzera italiana, Via G. Buffi 13, CH-6900 Lugano, Switzerland}
\affiliation{Faculty of Physics, University of Vienna, Boltzmanngasse 5, A-1090 Vienna, Austria}
\author{Stefan Wolf}
\affiliation{Faculty of Informatics, Universit\`a della Svizzera italiana, Via G. Buffi 13, CH-6900 Lugano, Switzerland}

\maketitle

\begin{abstract}
 \noindent 
One of the reasons for the heated debates around the interpretations of quantum theory is a simple confusion between the notions of formalism {\em versus\/} interpretation. 
In this note, we make a clear distinction between them and show that there are actually two \emph{inequivalent} quantum formalisms, namely the relative-state formalism and the standard formalism with the Born and measurement-update rules.
We further propose a different probability rule for the relative-state formalism and discuss how Wigner's-friend-type experiments could show the inequivalence with the standard formalism. The feasibility in principle of such experiments, however, remains an open question.
\end{abstract}
%
\section{Intoduction}
%
\subsection{Mathematical Formalisms and Interpretations}
\noindent
A physical theory\footnote{
We will use the terms \emph{scientific} and \emph{empirical} referring mostly to natural sciences and physics in particular.
}
comprises a \emph{mathematical formalism}, which allows for predicting the outcomes of scientific experiments, and some \emph{ontological interpretation}. In the case of quantum theory, the predictions are \emph{probabilistic} 
and often conflict with proposed descriptions of the experiment in terms of classical information \cite{bell1964einstein,kochen1975problem}. 
The arise of \emph{apparently classical} information during a measurement, \emph{i.e.}, \emph{a definite result}, poses a conceptual problem for quantum theory.
In what is called \emph{standard quantum mechanics}, the measurement-update rule, commonly associated with a collapse, is a break with the otherwise unitary evolution governed by the Schr\"odinger equation. The formalism, however, provides no indication about when to apply this rule: It does not state what qualifies some interactions as measurements but not others (\emph{the measurement problem}).\\
The \emph{relative-state formalism} \cite{everett1957relative} seems to avoid the problem by postulating universal, unitary quantum theory. This, however, detaches the formalism from predicting outcomes of experiments, for which some sort of Born rule \cite{born1954statistical} 
is needed. This is not specified by the original relative-state formalism at all, but the use of the Born rule has been motivated by a many-worlds interpretation and decision-theoretical arguments \cite{deutsch1999quantum,saunders2004derivation}. \\
We want to stress that universal, unitary quantum theory is a new type of \emph{formalism} which is fundamentally different from the measurement-update rule  of standard quantum mechanics. It is \emph{not} a new \emph{interpretation}; the many-worlds interpretation is the best-known interpretation of the relative-state formalism. One can regard a generalised version of Bohmian mechanics \cite{sudbery1986quantum} as a different interpretation of that formalism. \\
Throughout this note we will treat the relative-state formalism as a \emph{different formalism} than the Born and measurement-update rule of standard quantum mechanics. We postulate an alternative ``Born rule'' motivated by the work of G.~Hermann \cite{hermann1999foundations}. Equipped with this ``Born rule,'' the relative-state formalism reproduces the same probabilities as standard quantum mechanics for consecutive measurements on one quantum system --- the same level of observation. But the two formalisms are inequivalent in case of encapsulated observers --- different levels of observation, Wigner's-friend-type experiments \cite{wigner1963problem}. The latter was first considered by D. Deutsch \cite{deutsch1985quantum} and is made explicit in Chapter \ref{sec:WignerV}.

\subsection{Consistency of Quantum Theory}
\noindent
One manifestation of the quantum-measurement problem are Wigner's-friend-type \emph{gedankenexperiments}. In the original version, an \emph{observer} --- the friend $F$ --- performs a measurement $M_F$ on the quantum system emitted by the source $S$. Both the system and the friend (or the friend's memory) are then jointly measured by a \emph{superobserver} --- Wigner $W$ --- performing measurement $M_W$.
Standard quantum mechanics suggests that, according to the friend, the state of the system collapses to the eigenvector associated with the observed measurement result. To Wigner, however, the joint quantum system supposedly evolves unitarily.
Such a \emph{subjective-collapse} model, namely that each agent attributes a collapse merely to their own measurement, leads to \emph{seemingly contradictory predictions} among the agents, see~\cite{FrRen,baumann2016measurement}.
Descriptions of Wigner's-friend-type setups based on the relative-state formalism do not give rise to problematic predictions, neither do objective-collapse models, or any other version in which there is consensus on the application of the measurement-update rule.
The possibility of \emph{classical communication} between the agents in a Wigner's-friend-type experiment is essential for the \emph{problematic predictions} to give rise to an \emph{actual contradiction}, see Chapters \ref{sec:Terms_in_Science} and \ref{sec:Facts_of_the_World}.

\section{Scientific Theories}
\label{sec:Terms_in_Science}
\noindent
We regard a \emph{physical theory} as a mathematical formalism equipped with some ontological interpretation.
The formalism needs to predict the outcomes of scientific experiments, \emph{i.e.}, measurement results.
\begin{definition}\label{EQUI}
{\rm
Two mathematical formalisms are \emph{empirically equivalent} if they yield the same predictions for the outcomes of all possible experiments.
}
\end{definition}
\noindent
Examples for this are the general probabilistic theories formulation and operational quantum theory. They are designed to be empirically equivalent to standard quantum mechanics --- \emph{i.e.}, reproduce quantum probabilities --- and give insight into the general structure of the theory, see~\cite{hardy2001quantum,chiribella2010probabilistic,masanes2011derivation}.\\
We furthermore require that the predictions of measurement results render the theory \emph{falsifiable}, as proposed by K. Popper \cite{popper1935falsifizierbarkeit}. 
If a scientific theory gives rise to predictions that contradict the actual outcomes of experiments, or make contradictory predictions regarding these outcomes, it should be dismissed. 
We want to stress that falsification corresponds to a contradiction on the level of \emph{classical information}. Results of scientific experiments and statements regarding these experiments can be represented as entities of classical information, \emph{i.e.}, bit strings. If two such pieces of classical information are contradictory, they falsify the theory used to derive them. 
\begin{definition}\label{CI}
{\rm
Pieces of information are \emph{classical} if and only if they satisfy the requirements of \emph{interoperability} --- \emph{i.e.}, they can be copied --- and \emph{distinguishability} --- \emph{i.e.}, different information can be told apart perfectly.
}
\end{definition}
\noindent
Our definition of classicality throughout this note focuses on qualitative notions regarding information and not on the physical realisation. 
\begin{definition}\label{SCON}
{\rm
A \emph{scientific contradiction} is given by two pieces of contradictory classical information in one point in space and time.
}
\end{definition}
\noindent
Many quantum scenarios involving post-selection, for example \cite{aharonov2013quantum,wheeler1984quantum}, seem contradictory. But these apparent contradictions usually arise from attributing definite properties to systems at times, where these attributes do \emph{not} represent classical information.\\
\emph{Statements} about quantum experiments, furthermore, depend on the \emph{interpretation} of the quantum formalism. We argue that there are actually two \emph{empirically inequivalent} quantum formalisms, with different interpretations. If a certain \emph{combination} of formalism and ontological interpretation leads to a scientific contradiction, this combination should be excluded. \\
We want to emphasise the importance of linking any quantum formalism to classical information and measurement outcomes in order to have a physical theory as described above. 

\section{Two Quantum Formalisms}
\label{sec:Relative_State}
\noindent
Using our definitions in Chapter \ref{sec:Terms_in_Science}, we propose that there are \emph{two different quantum formalisms} for describing a measurement, which are \emph{empirically inequivalent} when considering \emph{encapsulated observers}, \emph{i.e.}, Wigner's-friend-type experiments. We will focus on pure states and projective measurements in the main text; a discussion of the general case can be found in Appendix \ref{App:General}. 
\subsection{Standard Quantum Mechanics}
\label{ssec:Copenhagen}
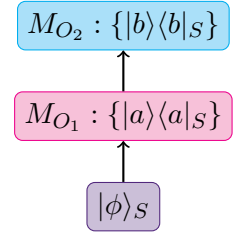
\begin{figure}
\centering
\begin{tikzpicture}[scale=0.6]
\node[fill=qcol!30, draw=qcol,rounded corners=3pt](s) at (0,0) {$\ket{\phi}_S$};
\node[fill=magenta!30,draw=magenta,rounded corners=3pt](M1) at (0,2) {$M_{O_1}: \{ \proj{a}_S \}$};
\node[fill=cyan!30, draw=cyan,rounded corners=3pt] (M2) at (0,4){$M_{O_2}: \{ \proj{b}_{S}\}$};

\draw[->,thick](s) --(M1);
\draw[->,thick](M1)-- (M2);
\end{tikzpicture}
\caption{
The same level of observation: Two observers $O_1$ and $O_2$ are performing consecutive measurements on a quantum system, initially in state $\ket{\phi} \in \H_S$. In this case, the measurement-update rule and the relative-state formalism give the same conditional probabilities for the measurement results of the two observers.
}
\label{pic:same_level}
\end{figure}

\noindent
In what is usually called \emph{standard quantum mechanics}, the description of a measurement is governed by two expressions. The \emph{Born rule} gives the probability of a measurement result for a quantum state  $\ket{\phi}$ and an observable $A=\sum_a a \proj{a}$,
\begin{equation} \label{QexpP}
p_{\phi} (a) =\tr(\proj{a}\proj{\phi}) =  |\braket{a}{\phi} |^2.
\end{equation}
The \emph{measurement-update rule} gives the quantum state of the system after the measurement,
\begin{equation} \label{pmeasurementP}
\ket{\phi}  \xrightarrow[\text{result: }a]{A} \ket{a}.
\end{equation}
Both the Born and the measurement-update rules can be motivated from the process-matrix formalism 
\cite{oreshkov2012quantum,chiribella2008quantum} by considering separate and consecutive measurements, see~\cite{shrapnel2018updating}. In particular, the measurement-update rule can be regarded as necessary for giving the correct probabilities for the results of consecutive measurements on one quantum system.\\ 
%
We will examine conditional probabilities of results of consecutive measurements by different observers. Consider two observers $O_1$ and $O_2$ successively measuring a quantum system initially in state $\ket{\phi}$, see Fig.\,\ref{pic:same_level}. Their measurements are given by $M_{O_1}: \{\proj{a}\}$ and $M_{O_2}: \{\proj{b}\}$. According to standard quantum mechanics, \emph{i.e.}, using equations \eqref{pmeasurementP} and \eqref{QexpP}, the conditional probability of result $b$ given $a$ is
\begin{equation} \label{coP-st}
\begin{aligned}
p_{\phi}(b|a) &= \frac{p_{\phi}(a,b)}{\sum_b p_{\phi}(a,b)} = \tr(\proj{b} \proj{a}) \\
&= |\braket{b}{a}|^2,
\end{aligned}
\end{equation}
where $p_{\phi}(a,b)=\tr(\proj{b}\proj{a}\proj{\phi}\proj{a}\proj{b})$ is the joint probability of $a$ and $b$. 
%
\subsection{The Relative-State Formalism}
\label{ssec:Everett_wigners_friend}
\noindent
In his thesis, H. Everett  III proposed the formalism of universal, unitary quantum mechanics, where the measurement is an entangling unitary between the quantum system and the observer.
Equivalently, measurements can be represented as isometries, correlating the (memory) state of the observer with the state of the observed system.
Consider a system in state $\ket{\phi}$ and an observer~$O$ performing a projective measurement $M_{O}: \{ \proj{a}_S \}$. 
\begin{IEEEeqnarray}{RCL}\label{eqn:stdIsom}
  V_{O} : \quad \H_S & \to &  \H_S \otimes \H_{O}\\ 
  \ket{a}_S & \mapsto & \ket{a}_S \otimes \ket{A_a}_O \quad \forall a \nonumber, 
\end{IEEEeqnarray}
where $\{\ket{A_a}\}_a$ is an orthogonal set recording the result. One can regard $\ket{A_a}$ as the state of the observer having seen result $a$.\\
A general state $\ket{\phi}$ then evolves as follows:
\begin{equation}
\label{pmeasurementE}
\ket{\phi}= \sum_{a} \braket{a}{\phi} \ket{a}  \mapsto  \sum_{a} \braket{a}{\phi} \ket{a} \otimes \ket{A_a} = \ket{\phi_{tot}}
\end{equation}
An overall entangled state, however, does not predict the outcomes of the measurement. Hence, there have been various attempts to justify the Born rule, Eq.\,\eqref{QexpP}, within the relative-state formalism. Independently of Everett's mathematical formulation, G.~Hermann~\cite{hermann1935naturphilosophischen} argued that the quantum state of the observed system is relative to the observing system and this relation is defined by and accessible only \emph{after} the measurement. It is the state of the observing system that represents the measurement result and with respect to which the state of the observed system is defined. We, therefore, propose the following ``Born rule'' for the evolution in Eq.\,\eqref{eqn:stdIsom}.
\begin{Postulate}\label{prop:EBorn1}
{\rm
Given a quantum state $\ket{\phi}$ and a measurement with outcomes $\{ a \}$, the probability of observing~$a$ is given by the trace of the projection onto the state~$\ket{A_a}$ of the observer having seen $a$ acting on the overall state evolving according to the relative-state formalism
}
\begin{equation}
q_{\phi}(a)= \tr(\mathds{1}_S\otimes \proj{A_a}_O \cdot V_O \proj{\phi} V_O^{\dagger}) \; .
\end{equation}
\label{eq:EBorn}
\end{Postulate}

\noindent
For one observer measuring a quantum system this is equivalent to the Born rule of standard quantum mechanics
\begin{equation}\label{P_a}
q_{\phi}(a)= |\braket{a}{\phi} |^2 = p_{\phi}(a)\; .
\end{equation}
Eq.\,\eqref{eq:EBorn}, however, can also be used to motivate the measurement-update rule in Eq.\,\eqref{pmeasurementP}.
\begin{Postulate}\label{prop:EBorn2}
{\rm
The joint probability of observations of states is given by the trace of the tensor product of the projectors onto those states acting on the overall state evolving according to the relative-state formalism.
}
\end{Postulate}
\noindent
Hence, the joint probability of observing the system in an eigenstate $\ket{a'}$ of the observer's measurement and the observer having seen outcome $a$ is
\begin{equation}\label{P(aA)}
\begin{aligned}
q_{\phi}(\ket{a'},a) &= \tr(\proj{a'}_S\otimes \proj{A_a}_O V_O \proj{\phi} V_O^{\dagger}) \\
&= |\braket{a}{\phi} |^2 \delta_{a,a'}. \\
\end{aligned}
\end{equation}
This can be read as the probability of the system being in the state $\ket{a}$, given that the observer has seen outcome $a$, is $1$, which is exactly the statement of Eq.\,\eqref{pmeasurementP}.\\
Moreover, let observer $O_1$ perform measurement $M_{O_1}:~\{\proj{a}\}$ on a quantum system in state~$\ket{\phi}$ and observer~$O_2$ then measure the same system, with~$M_{O_2}:~\{\proj{b}\}$.
According to Postulates~\ref{prop:EBorn1} and \ref{prop:EBorn2}, the conditional probability for the measurement results~$b$ given $a$ is
\begin{IEEEeqnarray}{RCL} \label{coP-ev}
&q_{\phi}(b|a) = \frac{q_{\phi}(a,b)}{q^M_{\phi}(a)} = \\
&\frac{\tr(\mathds{1}_S \otimes \proj{A_a}\otimes\proj{B_b} \proj{\Phi_{tot}})}
{\sum_b\tr(\mathds{1}_S \otimes\proj{A_a}\otimes\proj{B_b} \proj{\Phi_{tot}})}, \nonumber
\end{IEEEeqnarray}
where $\ket{\Phi_{tot}}=V_{O_2} V_{O_1} \ket{\phi}_S$.
Straight-forward calculations show that 
\begin{equation}\label{eqi_B_EB}
q_{\phi}(b|a)= |\braket{b}{a}|^2 = p(b|a).
\end{equation}
The conditional probabilities are the same as those given by the measurement-update rule of standard quantum mechanics. Hence, Eqs.\,\eqref{P_a} and \eqref{eqi_B_EB} justify the use the standard Born and measurement-update rules within the relative-state formalism as long as one considers the same level of observation. This agrees with former motivations of the standard Born rule for the relative-state formalism but is independent of the interpretation.
%
\subsection{Inequivalence}
\label{ssec:Comparison}
\begin{figure}
\centering
\begin{tikzpicture}[scale=0.7]
\draw[thick] (-2.5,-1) rectangle (2.5,3);

\node[fill=qcol!30, draw=qcol,rounded corners=3pt](s) at (0,0) {$\ket{\phi}_S$};
\node[fill=magenta!30,draw=magenta,rounded corners=3pt](M1) at (0,2) {$M_{O}: \{ \proj{a}_S \}$};
\node[fill=cyan!30, draw=cyan,rounded corners=3pt] (M2) at (6,1){$M_{SO}: \{ \proj{b}_{S,O}\}$};

\draw[->,thick](s) --(M1);
\draw[->,thick](2.5,1)-- (M2);
\end{tikzpicture}
\caption{
Different levels of observation: Observer $O$ performs a measurement on a quantum system $S$ in state $\ket{\phi}$. A superobserver $SO$ then performs a measurement on the joint system including the observer $O$.
For this type of consecutive measurements, the standard quantum formalism and the relative-state formalism yield different conditional probabilities for the measurement results. Hence, the two formalisms are \emph{empirically inequivalent}.
}
\label{pic:dif_level}
\end{figure}
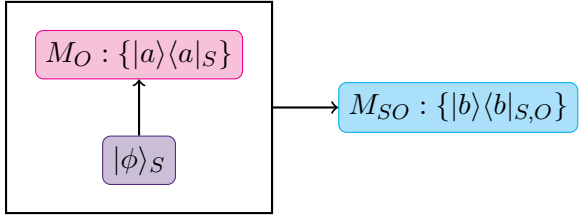

\noindent
So far the standard quantum and the relative-state formalisms yield the same probabilistic predictions. We will now consider encapsulated observers, see Fig.\,\ref{pic:dif_level}, and show that a particular version of standard quantum mechanics and the relative-state formalism equipped with the ``Born rule,'' Eq.\,\eqref{eq:EBorn}, are \emph{empirically inequivalent}.\\
Consider a quantum system in state $\ket{\phi}$, an observer $O$ measuring $M_{O}: \{ \proj{a}_S \}$ and a superobserver $SO$ measuring the joint system with $M_{SO}: \{ \proj{b}_{S,O}\}$. The standard Born rule, the measurement-update rule, and the observer measuring outcome $a$ are supposed to lead to the overall state
\begin{equation}\label{collapseW}
\ket{\phi}_s  \xrightarrow[\text{result: }a]{A} \ket{a}_S \otimes \ket{A_a}_S =\ket{a\otimes A_a}_{S,O} \; .
\end{equation}
For the situation depicted in Fig.\,\ref{pic:dif_level}, the conditional probabilities for the results $b$ measured by the superobserver are
\begin{equation}\label{collapseWcond}
p_{\phi}(b|a) =\frac{p_{\phi}(a,b)}{\sum_b p_{\phi}(a,b)}= | \braket{b}{a \otimes A_a}_{S,O}|^2 \;.
\end{equation}
This description represents an actual collapse of the wavefunction on the level of the observer.
%
According to the relative-state formalism, however, the observer's measurement corresponds to  an isometry, Eq.\,\eqref{eqn:stdIsom}. The same is true for the superobserver
\begin{IEEEeqnarray}{RCL}\label{eqn:IsomSO}
  V_{SO} : \H_S \otimes \H_{O} & \to &  \H_S \otimes \H_{O} \otimes \H_{SO}  \\ 
  \ket{b}_{S,O} & \mapsto & \ket{b}_{S,O} \otimes \ket{B_b}_{SO} ,\quad \forall b \nonumber. 
\end{IEEEeqnarray}
Let, now, $\{\ket{b}_{S,O}\}$ be a general orthonormal basis on the joint space $ \H_S \otimes \H_{O}$. The conditional probabilities of the results $b$ of the superobserver, given result $a$ of the observer, are
\begin{IEEEeqnarray}{RCL} \label{coP-EV}
&q_{\phi}(b|a) =\frac{\tr(\mathds{1}_S \otimes \proj{A_a}\otimes\proj{B_b} \proj{\Phi_{tot}})}{q^M_{\phi}(a)} \nonumber \\  \\
&=\frac{p_{\phi}(b|a)\mathlarger{\sum}\limits_{a' a''} \langle \phi |a''\rangle\langle a'|\phi \rangle 
\langle b | a' \otimes A_{a'}\rangle\langle a'' \otimes A_{a''}| b\rangle}{q^M_{\phi}(a)} \;. \nonumber 
\end{IEEEeqnarray}
This is equal to Eq.\,\eqref{collapseWcond} if and only if $\forall b:\exists a_0, a_1:\ket{b}_{S,O}= \ket{a_0}_S \otimes \ket{A_{a_1}}_O$. In that case the sum in Eq.\,\eqref{coP-EV} reduces to~$p_{\phi}(a)\delta_{a,a_0}$ and $q^M_{\phi}(a)=p_{\phi}(a)$,  hence,  $q_{\phi}(b|a)=p_{\phi}(b|a)$. \\
The empirical inequivalence between the two formalisms and its possible consequences for quantum theory become manifest in Wigner's-friend-type experiments. There, the observer and the superobserver are supposed to be thinking (or computing) entities knowing the setup and are, in principle, able to calculate the conditional probabilities for the measurements involved.

\section{Wigner's Friend}
\label{sec:WignerV}
\begin{figure}[h]
\centering
\vspace{-1em}
\begin{tikzpicture}[scale=0.6]
  \draw[thick, fill=gray!30] (-2,-2) rectangle (5.3,3);
  \node[fill=qcol!30, draw=qcol, thick] (s) at (-1,0) {$S$};
  \node[fill=magenta!30, draw=magenta, thick] (m) at (2,0) {$M_F$};
  \node[fill=magenta!20, draw=magenta!80, thick, rounded corners=3pt] (r1) at (4,1) {$\ket{\ua}\ket{U}$};
  \node[fill=magenta!20, draw=magenta!80, thick, rounded corners=3pt] (r2) at (4,-1) {$\ket{\da}\ket{D}$};
  \node[fill=cyan!30, draw=cyan, thick] (M) at (7.5,0) {$M_W$};
  \node[fill=cyan!20, draw=cyan!80, thick, rounded corners=3pt] (x1) at (10,1.5) {$\ket{\phi^+}\ket{+}$};
  \node[fill=cyan!20, draw=cyan!80, thick, rounded corners=3pt] (x2) at (10,-1.5) {$\ket{\phi^-}\ket{-}$};
  \draw[->] (s) -- node [midway,above] {$\ket{\phi}$} (m);
  \draw[->,shorten >=2pt,shorten <=2pt] (m) to[bend left] (r1);
  \draw[->,shorten >=2pt,shorten <=2pt] (m) to[bend right] (r2);

  \draw[->,thick] (5.3,0) to (M);
   \draw[->,shorten >=2pt,shorten <=2pt, thick] (M) to[bend left] (x1);
  \draw[->,shorten >=2pt,shorten <=2pt, thick] (M) to[bend right] (x2);
  
  \draw[<->,thick] (3.25,2.25) --node[midway, fill=qcol!30, draw=qcol, thick]{\larger{$C$}}  (7.25,2.25);
 \end{tikzpicture}
\caption{The source~$S$ emits a qubit $\ket{\phi}= 1/\sqrt{2}$ $(\ket{\ua} + \ket{\da})$, which is measured by the friend $F$ in the basis \{$\ket{\ua},$ $\ket{\da}$\}. Wigner $W$ then measures the joint system  in the superposition basis \{$\ket{\phi^{\pm}}= 1/\sqrt{2}(\ket{\ua\otimes U}\pm \ket{\da\otimes D})$\}. If $F$ applies the measurement-update rule for his measurement, while $W$ describes it via a unitary evolution, they will calculate different conditional probabilities for their results. This can be posed as contradicting statements  regarding $W$'s measurement result. These statements, however, only give rise to a \emph{scientific contradiction} if $F$ and $W$ can communicate classically ---\emph{i.e.}, their statements can be compared at some point.
}
\label{fig:Wigner}
\end{figure}
\begin{figure}
\centering
\includegraphics[width=0.9\linewidth]{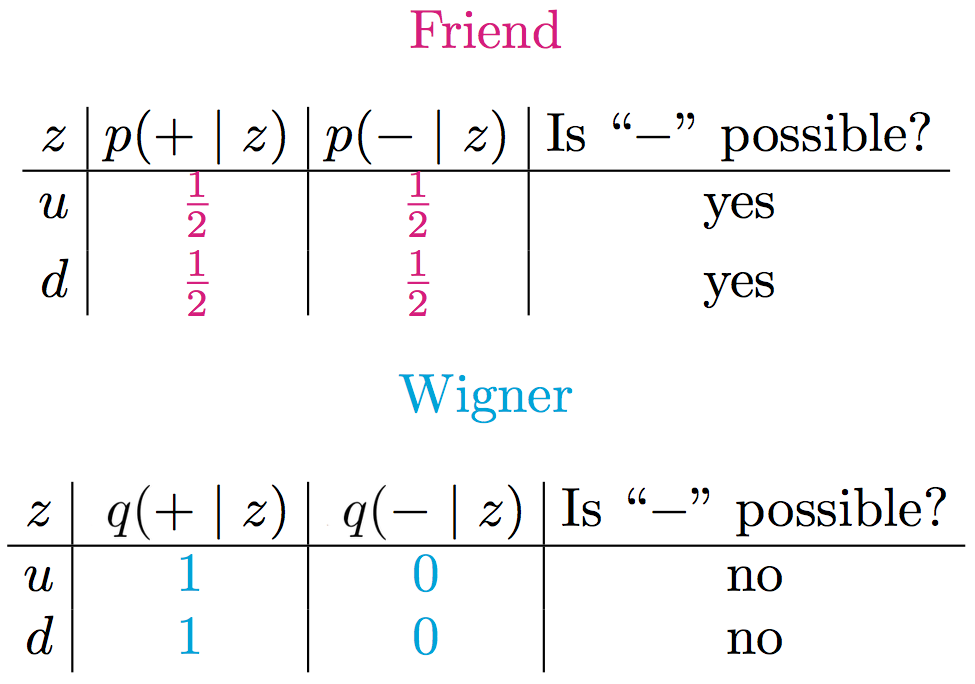}
\caption{
Let the source emit $\ket{\phi}= 1/\sqrt{2}(\ket{\ua}+\ket{\da})$, $F$ measure it in basis \{$\ket{\ua}, \ket{\da}$\}) and $W$ measure the joint system in the basis (e.g., \{$\ket{\phi^{\pm}}= 1/\sqrt{2}(\ket{\ua \otimes U}\pm$ $\ket{\da \otimes D})$\}. If the friend calculates the conditional probabilities with a collapse model, while Wigner uses the relative-state formalism, they will give contradicting answers to the question ``\emph{Can} $W$ measure ``$-$''?'' regardless of $F$'s measurement result.
}
\label{Contra1}
\end{figure}

\noindent
The simplest version of the  Wigner's-friend experiment is depicted in Fig.\,\ref{fig:Wigner}. A qubit in state $\ket{\phi}$ is measured by observer $F$ after which a superobserver~$W$ measures the joint system of the qubit and the observer's memory. The superobserver is said to have \emph{full quantum control}.
There are two unproblematic descriptions of the situation. Both agents apply the measurement-update rule for every measurement and, hence, calculate the conditional probabilities for their results according to Eq.\,\eqref{collapseWcond}. This case corresponds to an \emph{objective collapse} during a measurement. Alternatively, all agents can use the relative-state formalism and calculate the conditional probabilities using Eq.\,\eqref{coP-EV}. We might call this the \emph{no-collapse} model.
It has been argued, however, that standard quantum mechanics suggests that $F$ uses the measurement-update rule after his measurement and, therefore, Eq.\,\eqref{collapseWcond}. But~$W$, to whom the joint quantum system evolves unitarily, uses Eq.\,\eqref{coP-EV} instead. We call this the \emph{subjective-collapse} model.\\
In the case of qubit state $\ket{\phi}= 1/\sqrt{2}(\ket{\ua}+\ket{\da})$, $F$ measuring in the $\sigma_z$-basis, \{$\ket{\ua}, \ket{\da}$\}, and $W$ measuring in basis  
\{$\ket{b}_{S,O}$\}, where $\ket{\ua\otimes U}=\alpha \ket{b_1} + \beta \ket{b_2}$ and $\ket{\da \otimes D}=\beta \ket{b_1} - \alpha \ket{b_2}$, the two formalisms yield the following conditional probabilities.
\begin{center}
{\color{magenta}measurement-update rule}
\begin{equation}\label{mutab}
  \begin{array}{c|c|c}
   z & p( b_1\mid z) & p( b_2\mid z) \\ \hline
   u & \color{magenta}\alpha^2 & \color{magenta}\beta^2  \\
   d & \color{magenta}\beta^2  & \color{magenta}\alpha^2
  \end{array}
\end{equation}

{\color{cyan}relative-state formalism}
\begin{equation}\label{rstab}
  \begin{array}{c|c|c}
   z & q( b_1\mid z) & q( b_2\mid z) \\ \hline
   u & \color{cyan}\left(\frac{\alpha+\beta}{2\alpha}\right)^2 &\color{cyan}\left(\frac{\alpha-\beta}{2\beta}\right)^2  \\
   d & \color{cyan}\left(\frac{\alpha+\beta}{2\beta}\right)^2  & \color{cyan}\left(\frac{\alpha-\beta}{2\alpha}\right)^2
  \end{array}
\end{equation}
\end{center}
For Wigner measuring in the basis \{$\ket{\phi^{\pm}}= 1/\sqrt{2}(\ket{\ua\otimes U}\pm \ket{\da\otimes D})$\}, this leads to contradicting statements of $F$ and $W$ regarding the possibility of $W$ measuring result``$ -$,'' see Fig.\,\ref{Contra1}.

\begin{figure}[h]
\centering
\begin{flushleft}
a
\end{flushleft}
\vspace{-1.5em}
\begin{tikzpicture}[scale=0.6]
  \draw[thick] (-1,-1) rectangle (4.3,1);
  \node[fill=qcol!30, draw=qcol, thick] (c) at (0,0) {$C$};
  \node[fill=magenta!30, draw=magenta, thick] (mf1) at (3,0) {$M_{F1}$};
  \node[fill=magenta!30, draw=magenta, thick] (MA) at (6,0) {$M_A$};
  \draw[->] (c) -- node [pos=0.5,above] {$\ket{\phi_C}$} (mf1);
  \draw[->,thick] (4.3,0) to (MA);
  \begin{scope}[shift={(2,-3)}]
  \draw[thick] (-1,-1) rectangle (4.3,1);
  \node[fill=qcol!30, draw=qcol, thick] (s) at (0,0) {$S$};
  \node[fill=cyan!30, draw=cyan, thick] (mf2) at (3,0) {$M_{F2}$};
  \node[fill=cyan!30, draw=cyan, thick] (MW) at (6,0) {$M_W$};
  \draw[->] (s) --node [pos=0.5,above] {$\ket{\phi_S}$} (mf2);
  \draw[->,thick] (4.3,0) to (MW);
  \end{scope}
  \draw[->,thick] (mf1) to (s);
\end{tikzpicture}\\
\begin{flushleft}
b
\end{flushleft}
\vspace{-1.5em}
\includegraphics[width=0.9\linewidth]{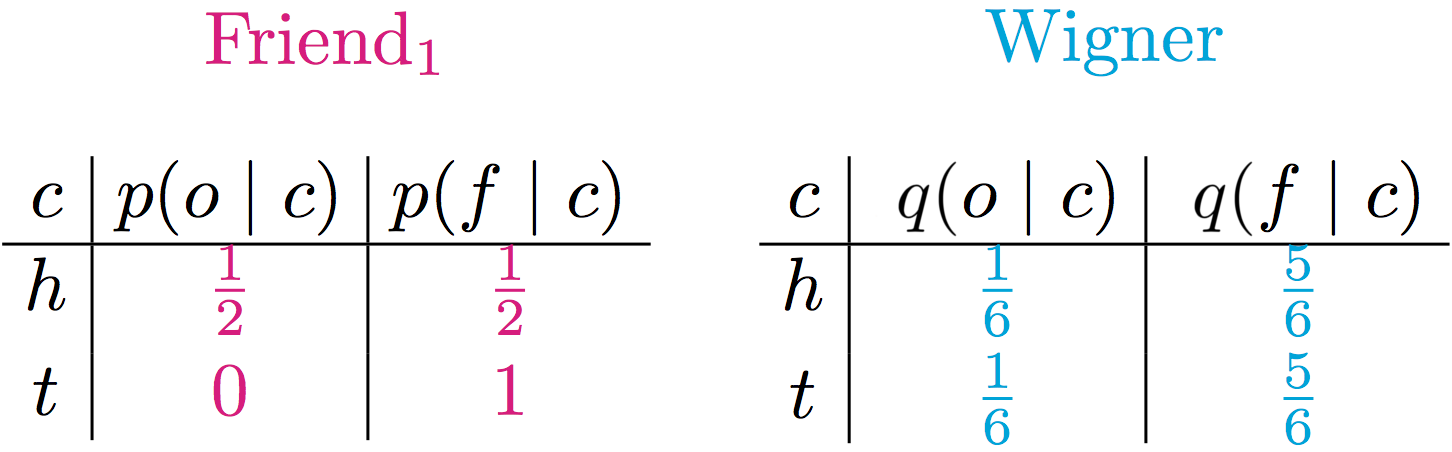}
\caption{a. The source in the first laboratory emits a coin state $\ket{\phi_C}$ which is measured by $F_1$, with outcomes \{$h,t$\}. Depending on the result, $F_1$ sends state $\ket{\phi_S}$ which is measured by $F_2$. Assistant $A$ and Wigner~$W$ perform measurements in superposition bases, with outcomes \{$f,o$\}, on the first and second joint system, respectively.
b. If $F_1$ uses the measurement-update rule, while $W$ describes the  measurement according to the relative-state formalism, they get different conditional probabilities for their outcomes. Upon observing outcome $t$,  $F_1$ will predict that $W$ will measure $f$ with certainty. According to $W$, however, there should be instantiations of the experiment where $F_1$ observes $t$ but he will measure $o$. }
\label{fig:modWigner}
\end{figure}

\noindent
In~\cite{FrRen} the authors derive another contradiction for a combination of two standard Wigner's-friend experiments, see Fig \ref{fig:modWigner}.
The four parties involved are two observers --- friend $F_1$ and friend $F_2$ --- and two superobservers --- Wigner $W$ and his assistant $A$. This contradiction is more striking since it can be phrased in terms of \emph{deterministic} predictions of the results in \emph{one} particular run of the experiment. The arise of $F_1$'s prediction, however, is empirically equivalent to him 
using the state~$\ket{t}_C\otimes \ket{T}_{F_1}\otimes \ket{\rightarrow}_S$ to calculate conditional probabilities according to Eq.\,\eqref{collapseWcond}, where the joint probability is
\begin {equation*}
\begin{aligned}
&p_{\phi_C}(c,w)=\\
&\tr(K_w \proj{C} (\proj{\phi_C}\otimes \mathds{1})\proj{C} K_w^{\dagger}), 
\end{aligned}
\end{equation*}
with~$\ket{C}=\ket{c}_C\otimes \ket{C_c}_{F_1}\otimes \ket{\phi_S(c)}_S$ and~$K_w=\proj{w}_{F_2,S}V_{F_2}$. $W$'s predictions, on the other hand, are empirically equivalent to using the overall entangled state arising from the unitary evolution and Eq.\,\eqref{coP-EV}.\\
In general, the subjective-collapse model presents a problematic description of 
experiments featuring encapsulated observers. Whether it should be excluded depends on whether
it gives rise to a scientific contradiction, as discussed in Chapters~\ref{sec:Facts_of_the_World} and \ref{sec:PWigner}. \\
Moreover, the Wigner's-friend experiment can distinguish between the two quantum formalisms --- \emph{i.e.,} Born and (objective) measurement-update rule or the relative-state formalism with the ``Born rule''--- when they are applied for all measurements  by all agents involved. 

\section{Facts of the World and Interpretations}
\label{sec:Facts_of_the_World}
\noindent
Interpretations of scientific theories attribute ontological concepts to the entities of the mathematical formalism and can roughly be regarded as answering the question ``What does the formalism mean?''. 
In general this specifies a connection between the formalism and some notion of reality, \emph{i.e.}, \emph{facts}. 
\begin{definition}\label{INT}
{\rm
If two \emph{scientific theories} are empirically equivalent, they are called different \emph{interpretations} of the same \emph{formalism}.
}
\end{definition}
\noindent
Note that, according to our Defs.\,\ref{EQUI} and \ref{INT}, only a formalism together with an interpretation constitutes a scientific theory\footnote{
The interpretations of quantum theory presented here are by no means a complete list, but rather serve to illustrate our argument. 
}. We want to point out, however, that ``Shut up and calculate!'' also involves an interpretation of the quantum formalism, namely that the formalism is merely a tool 
to calculate probabilities for measurement outcomes, which are the only facts in this approach. \\
%
If one assumes the quantum formalism to describe an \emph{objective reality} one might talk about \emph{facts  per se}. In that case, all agents in a Wigner's-friend-type experiment must use either the collapse formalism or the relative-state formalism (for all measurements) or the theory is self-contradictory. \\
%
Interpretations of this type are for example, part of all objective collapse models, the many-worlds interpretation and Bohmian mechanics. A modification of the standard quantum formalism as in~\cite{ghirardi1986unified} allows for an objective interpretation with a ``real'' collapse. 
We argue that the many-worlds interpretation and generalised Bohmian mechanics are different interpretations of the relative-state formalism according to Def.\,\ref{INT}.
Both use the same unitary evolution of a \emph{global wave function} regardless of a measurement happening or not. In case of the many-worlds interpretation, this global wave function corresponds to the state of reality and a \emph{multiverse}, a branch of which represents what we observe. In Bohmian mechanics the global wave function is the \emph{pilot wave} but the state of reality is the so-called \emph{real state vector}. The latter arises from the decomposition of the pilot wave into vectors of viable subspaces, which is related to Bell's beables \cite{sudbery1986quantum, bell2004speakable}.
The real state vectors evolve probabilistically due to transitions between the viable subspaces. But since the probabilities of measurement results, via the real state vectors, are determined by the pilot wave, the predictions match those of the relative-state formalism, see Appendix \ref{App:Bohm} for an explicit example.\\
If one adopts a subjective application of the two formalisms, \emph{i.e.}, \emph{subjective collapse}, as a consequence there are only \emph{facts relative to the observer}, see~\cite{brukner2015quantum}. In other words, a subjective-collapse model requires a subjective interpretation like, for example, relational quantum mechanics or QBism \cite{rovelli1996relational,fuchs2010Qbism}. This circumvents consistency requirements imposed by the idea of the quantum formalism describing an objective reality per se. Note, however, that a \emph{scientific contradiction} in a Wigner's-friend-type experiment based on the subjective collapse would exclude that model irrespective of the interpretation.\\ 
To, once more, stress the distinction between a formalism and an interpretation, we want to point out that our reading of the relative-state formalism, and in particular our motivation for the ``Born rule,'' is along the lines of facts relative to the observer.

\section{Wigner's-Friend-Type Experiments and Classical Information}
\label{sec:PWigner}
\noindent
A \emph{scientific contradiction} comprises contradicting pieces of classical information and must not arise for a scientific theory. Considering classical information in Wigner's-friend-type experiments is essential for deciding, whether the subjective collapse model is scientifically contradictory. Even given a subjective interpretation, classical information should be intersubjectively agreed on by all agents and, therefore, must not be contradictory. In other words, a subjective interpretation cannot resolve the problematic descriptions in Chapter \ref{sec:WignerV} if they give rise to a scientific contradiction.\\
The notion of classicality in Def.\,\ref{CI} allows for incorporating it into the quantum formalism. An orthonormal basis \{$\ket{i}$\} represents classical information if and only if all \emph{accessible  observables} are diagonal in that basis. 
%
%
\noindent
This notion of classical information is independent of whether one attributes the arise of classicality to the dynamics of the observed system \cite{zurek2003decoherence} or the observation itself \cite{kofler2007classical}. \\ 
In the context of Wigner's-friend-type experiments, the observer's measurement result does not represent classical information, since it does not satisfy interoperability. 
This is reflected in the fact that the inconsistent predictions arise only if there is \emph{no classical record} revealing the observers' results. Consider a register system \{$\ket{x_i}\in \H_R$\} representing classical information. In the relative-state formalism, this corresponds to the isometry 
\begin{IEEEeqnarray}{RCL}\label{eqn:IsomC}
  V_{O} : \quad \H_S & \to &  \H_S \otimes \H_{O} \otimes \H_{R}\\ 
  \ket{a}_S & \mapsto & \ket{a}_S \otimes \ket{A_a}_O \otimes \ket{x_a}_R\quad \forall a \nonumber, 
\end{IEEEeqnarray}
where $\ket{x_i}\in \H_R$ can be thought to represent the observer's statements. Since the \{$\ket{x_i}$\} form a basis, the statements they encode should be informationally complete, by which we mean that $\forall s: \ket{x_j} \cong s\; \exists \ket{x_{j'}} \in \{ \ket{x_i}  \}: \ket{x_{j'}} \cong \lnot s$.\\
%
The conditional probability for the measurement results can then be defined as
\begin{IEEEeqnarray}{RCL} \label{coP-EVC}
&q_{class}(b|a) :=  \frac{q_{\phi}(a, x_a, b)}{\sum_b q_{\phi}(a, x_a, b)} \nonumber\\[0.5em]
&=\frac{\tr(\proj{x_a}\otimes \proj{A_a}\otimes\proj{B_b}\proj{\Phi_{tot}})}
{\tr(\proj{x_a}\otimes \proj{A_a} \proj{\Phi_{tot}})} \nonumber\\[0.5em]
&\Rightarrow q_{class}(b|a)\neq q_{\phi}(b|a).
\end{IEEEeqnarray}
It follows from this that the extended Wigner's-friend experiment in Fig.\,\ref{fig:modWigner} cannot give rise to a scientific contradiction: $F_1$'s deterministic prediction ``$w=f$'' arises only for $\ket{t}_C\otimes \ket{T}_{F_1}\otimes \ket{\rightarrow}_S$. If there is a classical record of this statement, the overall probability distribution matches the predictions of $F_1$:
\begin{center}
Extended Wigner's-Friend Experiment
\begin{equation}
  \begin{array}{c|c|c}
   c &{ \color{qcol}q_{class}}( o\mid c) & {\color{qcol}q_{class}}( f\mid c) \\ \hline
   h & \frac{1}{2} & \frac{1}{2}  \\[0.25em]
   t & 0  & 1,
  \end{array} 
\end{equation}
\end{center}
Since the prediction ``$p(+)=p(-)=\frac{1}{2}$'' in the original Wigner's-friend experiment depicted in Fig.  \ref{fig:Wigner} arises for both $\ket{\ua}_S\otimes \ket{U}_F$ and $\ket{\da}_S\otimes \ket{D}_F$ the existence of a classical record of this statement does not affect the conditional probabilities. 
\begin{center}
Original Wigner's-Friend Experiment
\begin{equation}\label{PclassEW}
  \begin{array}{c|c|c}
   z &{ \color{qcol}q_{class}}( + \mid z) & {\color{qcol}q_{class}}( - \mid z) \\ \hline
   u & 1 & 0  \\[0.25em]
   d & 1 & 0
  \end{array}
\end{equation}
\end{center}
It is, however, not clear to us whether the observer's computation leading to this statement is compatible with being under \emph{full quantum control}. 
The respective classical register would be~\{$\ket{s_i}$\} with $i=0,1,2$, where $\ket{s_0\cong ``\text{no measurement}"}$, $\ket{s_1\cong ~``50:50"}$ and $\ket{s_2 \cong ``\lnot (50:50)"}$. Note that the map 
\begin{equation*}
\ket{z \otimes Z}_{S,O}\otimes \ket{s_0}_R \mapsto \ket{z\otimes Z}_{S,O}\otimes\ket{s_1}_R, 
\end{equation*}
with $\ket{s_1}$ being the same for both $\ket{\ua \otimes U}$ and $\ket{\da \otimes D}$, cannot be a unitary on the joint space $\H_S \otimes \H_O \otimes \H_R$, where $\H_R= \text{span}\{\ket{s_i}\}$. 

\section{Discussion}
\label{sec:Discussion}
\noindent 
An onthological {\em interpretation\/} of a mathematical formalism establishes a correspondence between the mathematical objects of the theory and elements of some ontology. The latter is associated with \emph{factual} classical information. In most physical theories, an identification of the entities in the formalism with an ontological description in terms of classical information is usually unambiguous and generally agreed upon. In the case of quantum theory, however, attempts to do so usually lead to paradoxes, like \emph{Schr\"odinger's cat} or \emph{spooky action at a distance} to explain Bell-non-local correlations.
Therefore, there are various \emph{interpretations} of quantum theory ascribing different notions of a factual reality to different entities in the formalism. We defined different interpretations of the same formalism to be \emph{empirically equivalent}, \emph{i.e.,} they give the same predictions for the results of all experiments. We showed that there are two \emph{empirically inequivalent} quantum formalisms regarding the description of a measurement --- the standard Born- and measurement-update rule and the relative state formalism. The difference between the two becomes apparent only in Wigner's-friend-type experiments. There, the possibility of classical communication and the subjective application of the two formalisms might give rise to a \emph{scientific contradiction}, i.e., contradictory classical information.\\
We attempt in this note to contribute to the discussion on the meaning and interpretation
of quantum theory by making clearer the separation line between mathematical
formalisms and their ontological interpretations --- they have, in our opinion,
too often been confused.
\begin{acknowledgments}
\noindent
This work is supported by the Swiss National Science Foundation (SNF) and the NCCR QSIT. We thank \"Amin Baumeler, \v{C}aslav Brukner, Julien D. Degorre, Paul Erker, Arne Hansen and Renato Renner for helpful discussions. 
\end{acknowledgments}

\bibliography{Refs}

\onecolumn\newpage
\begin{appendix}
\section{General Formulation}
\label{App:General}
\noindent
Consider a general quantum state $\rho$ of some system and an observable \{$K_a^{\dagger} K_a$\}, where \{$a$\} are the outcomes and~$K_a$ is the Kraus operator associated with outcome~$a$. The probability of outcome $a$ is then
\begin{equation} \label{Qexp}
p_{\rho} (a) =\tr(K_a\rho K_a^{\dagger}). 
\end{equation}
The measurement-update rule for the general case is
\begin{equation} \label{pmeasurement}
\rho \xrightarrow[\text{result: }a]{}  \frac{1}{p_{\rho} (a)} K_a \rho K_a^{\dagger}.
\end{equation}
The action of a Kraus operator $K_a$ on $\rho$ can always be written as 
\begin{equation}\label{Kraus}
\rho \rightarrow \bra{a} U_{s,x} \left( \rho \otimes \proj{a_0}\right) U_{s,x}^{\dagger} \ket{a},
\end{equation}
where \{$\ket{a}$\} form an orthonormal basis in an ancillary space $\H_X$ and $U_{x,s}$ is a unitary operator on $\H_S \otimes \H_X$. Using representation \eqref{Kraus}, the measurement in the relative-state formalism is given by an isometry correlating the ancillary system with the observer's memory:
\begin{IEEEeqnarray}{RCL}\label{eqn:KIsom}
  V_{O} : \quad \H_X & \to &  \H_X \otimes \H_{O}\\ 
  \ket{a}_X & \mapsto & \ket{a}_X \otimes \ket{A_a}_O \quad \forall a \; , \nonumber  
\end{IEEEeqnarray}
and a general state  $\rho$ then evolves as follows:
\begin{IEEEeqnarray}{RCL}\label{RSrho}
\rho &\rightarrow& \rho_{tot}= \sum_{c c' a a'} 
\rho_{c c' a a'} \ket{c}\bra{c'}_S \otimes V_{O} \ket{a}\bra{a'}_X V_{O}^{\dagger}, \\
\text{with } & \rho_{c c' a a'}& 
=\bra{c} \otimes \bra{a} U_{s,x} \left( \rho \otimes \proj{a_0}\right) U_{s,x}^{\dagger} \ket{a'} \otimes \ket{c'}. \nonumber
\end{IEEEeqnarray}
Again, straight-forward (but tedious) calculation shows that
\begin{IEEEeqnarray}{RCL}\label{RSrho}
q_{\rho}(a)&=& \tr(\mathds{1}_{S,X}\otimes \proj{A_a}_O \cdot \rho_{tot})
= \sum_{c} \rho_{ccaa} = \tr(K_a\rho K_a^{\dagger})=p_{\rho}(a). \nonumber
\end{IEEEeqnarray}
For two observers performing consecutive measurements on one system --- 
$M_{O_1}: \{K_a^{\dagger} K_a \}$ and $M_{O_2}: \{K_b^{\dagger} K_b\}$ --- one gets
\begin{IEEEeqnarray}{RCL}\label{RSrho}
q_{\rho}(b|a)&=& \frac{1}{q_{\rho}(a)} \tr(\mathds{1}\otimes \proj{A_a}_{O_1} \otimes \proj{B_b}_{O_2}\cdot \rho'_{tot})=\frac{1}{q_{\rho}(a)} \sum_c \rho_{ccaabb} \nonumber \\
&=& \frac{1}{p_{\rho}(a)} \tr(K_b K_a \rho K_a^{\dagger} K_b^{\dagger}) \nonumber 
= \tr\left( K_b \left(\frac{K_a \rho K_a^{\dagger}}{p_{\rho}(a)}\right) K_b^{\dagger}\right) = p_{\rho}(b|a),
\end{IEEEeqnarray}
\begin{IEEEeqnarray}{RCL}\label{RSrho}
\text{with } \;\; &\rho'_{tot}&= \sum_{\substack{c c' a a'\\ b b'}} \rho_{c c' a a' b b'}
\ket{c}\bra{c'}\otimes V_{O_1}\ket{a}\bra{a'}V_{O_1}^{\dagger}
\otimes V_{O_2}\ket{b}\bra{b'}V_{O_2}^{\dagger},
\nonumber \\
\text{where } &\rho_{cc'aa'bb'}&=  \bra{cab} U_{s,y}(U_{s,x} \rho \otimes \proj{a_0}_X U_{s,x}^{\dagger})\otimes \proj{b_0}_Y U_{s,y}^{\dagger}\ket{c'a'b'} \nonumber
\end{IEEEeqnarray}
and $\ket{cab}=\ket{c}_S\otimes \ket{a}_X\otimes\ket{b}_Y$.
Hence, on the same level of observation, the relative-state formalism together with Postulates \ref{prop:EBorn1} and \ref{prop:EBorn2} gives the same probabilistic predictions as the standard Born  and measurement-update rules also for generalised measurements. 
One can motivate the use of the measurement-update rule, \eqref{pmeasurement}, within the relative-state formalism by Eq.~\eqref{RSrho}. If the conditional probabilities for consecutive measurements on one quantum system are calculated by the ``Born rule'', the measurement-update rule gives the correct state assignment when using the standard Born rule.

\section{Relative-State Formalism and Bohmian Mechanics} 
\label{App:Bohm}
\noindent
It has been show already, see for example~\cite{durr1996bohmian}, that for the same level of observation, Bohmian mechanics gives the same probabilistic predictions as the standard quantum formalism. \\
For different levels of observation we consider the example in Fig.\,\ref{fig:modWigner}. A description of the extended Wigner's-friend experiment in terms of generalised Bohmian mechanics was presented in~\cite{sudbery2017single}, including the following joint probabilities for $A$'s and $W$'s outcomes $(a,w)$ calculated for different agents:
\begin{equation}\label{pBohm}
  \begin{array}{l|c|c|c|c}
   p(a,w)& p( o,o) & p( o,f) &p(f,o) & p(f,f)\\ \hline
   F_1 & \frac{1}{12} & \frac{5}{12}& \frac{1}{12} & \frac{5}{12} \\[0.25em]
    F_2 & \frac{1}{12} & \frac{1}{12}& \frac{5}{12} & \frac{5}{12} \\[0.25em]
   A & \frac{1}{4} & \frac{1}{4}& \frac{1}{20} & \frac{9}{20} \\[0.25em]
   W & \frac{1}{12} &\frac{1}{12} &\frac{1}{12} &\frac{3}{4} 
  \end{array}
\end{equation}\\
The last line representing $W$ equals the joint probabilities~$q_{\phi}(a,w)$ calculated from the state evolving according to the relative-state formalism and Eq.\,\eqref{eq:EBorn}.
\begin{equation}\label{extWBohm}
\begin{aligned}
&q_{\phi}(a,w)=\tr(\mathds{1}\otimes \proj{A_a}\otimes\proj{W_w} \cdot \proj{\phi_{tot}})\\
&\text{with: } 
\ket{\phi_{tot}}=V_{W}V_{A}V_{F_2} V_{F_1} \proj{\phi_C}V_{F_1}^{\dagger}V_{F_2}^{\dagger}V_{A}^{\dagger}V_{W}^{\dagger} \, . \\
\end{aligned}
\end{equation}\\
The probability distribution of $A$ can be obtained by renormalising the conditional probabilities $q_{\phi}(w|a)$ calculated according to Eq.\,\eqref{coP-ev}. The friends' distributions arise when one takes the conditional probabilities $q_{\phi}(a|z)$ for $F_2$, who measures result $z$, and $q_{\phi}(w|c)$ for $F_1$, who measures result $c$, and renormalises them to give distribution for both results. One might think of this as each friend neglecting the measurement performed on them and predicting the superobservers' outcomes depending on their result.
We leave it for future work to show empirical equivalence for the general case but argue that the comparison above supports our claim that one can regard generalised Bohmian mechanics as an interpretation of the relative-state formalism according to Def.\,\ref{INT}.

%
\end{appendix}

\end{document}